\documentstyle[12pt]{article}
\begin{document}
\large
\begin{center}
{\bf Some Unsettled Questions in the Problem of Neutrino
Oscillations}
\par
\vspace{0.5cm}
Beshtoev Kh. M.
\par
\vspace{0.5cm}
JINR, Dubna\\
\end{center}
\vspace{0.71cm}
\par
{\bf  Abstract}\\

It is noted that the theory of neutrino oscillations can be
constructed only in the framework of the particle physics theory,
where is a mass shell conception and then transitions
(oscillations) between neutrinos with equal masses are real  and
between neutrinos  with different masses are virtual. It is
necessary to solve the question: which type of neutrino
transitions (oscillations) is realized in nature? There can be
three  types of neutrino transitions (oscillations).
At present  it is considered that Dirac and Majorana neutrino
oscillations can be realized. It is shown that we cannot put
Majorana neutrinos in the standard weak interactions theory without
violation of  the gauge invariance. Then it is obvious that there
can be only realized transitions (oscillations) between Dirac
neutrinos with different flowers. Also it is shown that the mechanism of
resonance enhancement of neutrino oscillations in matter cannot be
realized without violation of the law of energy-momentum
conservation. Though it is supposed that we see neutrino oscillations
in experiments, indeed there only transitions between neutrinos are registered.
In order to register neutrino oscillations it  is
necessary to see second or even higher neutrino oscillation modes
in experiments. For this purpose we can use the elliptic character
of the Earth orbit. The analysis shows that the SNO
experimental results do not confirm smallest of $\nu_e \to
\nu_{\tau}$ transition angle mixings, which was obtained in CHOOZ
experiment. It is also noted that there is contradiction between SNO,
Super-Kamiokande, Homestake and the SAGE and GNO (GALLEX) data.

\newpage

\par
{\bf I. Introduction}\\

\par
The suggestion that, by analogy with $K^{o},\bar K^{o}$
oscillations, there could be neutrino oscillations (i.e., that
there could be neutrino-antineutrino oscillations $\nu \rightarrow
\bar \nu$) was considered by Pontecorvo [1] in 1957. It was
subsequently considered by Maki et al. [2] and Pontecorvo [3] that
there could be mixings (and oscillation) of neutrinos of different
aromas (i.e., $\nu _{e} \rightarrow \nu_{\mu }$ transitions).
\par
The problem of solar neutrinos arose after the first experiment has been
performed in order to measure the flux of neutrinos from the Sun by the
$^{37}Cl - ^{37}Ar$ [4] method. The flux was found to be several
times smaller than expected from calculations made in accordance
with the standard solar model (SSM) [5]. It was suggested in [6]
that the solar neutrino deficit could be explained by Majorana neutrino
oscillations. Subsequently, when the result of the experiment at
Kamiokande [7] confirmed the existence of the deficit relative to
the SSM calculations, one of the attractive approaches to the
explanation of the solar neutrino deficit became resonant
enhancement of neutrino oscillations in matter [8]. Resonant
enhancement of neutrino oscillations in matter was obtained from
Wolfenstein's equation for neutrinos in matter [9]. It was noted
in Ref. [10] that Wolfenstein's equation for neutrinos in matter
is an equation for neutrinos in matter in which they interact with
matter not through the weak but through a hypothetical weak
interaction that is left-right symmetric. Since only left components of
neutrinos participate in the standard weak interactions,
the results obtained from Wolfenstein's equation have no direct
relation to real neutrinos.
\par
Later experimentalists obtained the first results on the Gran
Sasso $^{71}Ga - ^{71}Ge$  experiment [11], that within a $3\sigma
$ limit did not disagree with the SSM calculations. The new data
from the SAGE experiment [12] is fairly close to the Gran Sasso
results.
\par
After the discovery of neutrino transitions (oscillations) on
Super-Kamiokande [13, 14] and on SNO [15], it is necessary to analyze
the situation, which arises in the problem of neutrino oscillations.
\par
This work is devoted to discussion of some theoretical and experimental
questions, which have been kept unsettled in the problem of neutrino
oscillations. These are: construction of the correct neutrino
oscillations theory;  quest for the type of neutrino transitions
(oscillations); determination of  the concrete mechanism of
neutrino transitions; real observation of the neutrino
oscillations on experiments; solution of some contradictions
 in experiments (CHOOZ problem, inconsistency SNO, SK, Homestake and
GNO, SAGE data).\\

\par
{\bf II. THEORY}\\

{\bf 1. The Theory of Neutrino Oscillations}\\

\par
In the old theory of neutrino oscillations [6, 16], constructed in
the framework of Quantum Mechanics in analogy with the theory of
$K^{o}, \bar{K}^{o}$ oscillation, it is supposed that mass
eigenstates are $\nu_{1}, \nu_{2}, \nu_{3}$ neutrino states but
not physical neutrino states $\nu_{e}, \nu_{\mu }, \nu_{\tau}$,
and that the neutrinos $\nu_{e}, \nu_{\mu }, \nu_{\tau}$ are
created as superpositions of $\nu_{1}, \nu_{2}, \nu_{3}$  states.
This means that the $\nu_{e}, \nu_{\mu }, \nu_{\tau}$ neutrinos
have no definite mass, i.e. their masses may vary in dependence on
the  $\nu_{1}, \nu_{2}, \nu_{3}$ admixture in the $\nu_{e},
\nu_{\mu }, \nu_{\tau}$  states. On the example of $K^o,  \bar
K^o$ mesons (eigenstates of the strong  interactions) we can well
see that on the time $10^{-23} sec$ (a typical time of the strong
interactions) the $K^o_1, K^o_2$ mesons-eigenstates of the weak
interactions cannot be created since its typical time is
$10^{-6}-10^{-8} sec$. Naturally, in this case the law of
conservation of the energy and the momentum of the neutrinos is
not fulfilled. Besides, every particle must be created on its mass
shell and it will be left on its mass shell while passing through
vacuum. It is clear that this picture is incorrect.
\par
In the modern theory on neutrino oscillations [17]-[18],
constructed in the framework of the particle physics theory, it is
supposed that:\\

\par
1)  The  physical  states  of  the $\nu_{e}, \nu_{\mu },
\nu_{\tau}$ neutrinos   are eigenstates of the weak interaction
and, naturally, the mass matrix of $\nu_{e}, \nu_{\mu },
\nu_{\tau}$ neutrinos is diagonal. All  the  available
experimental results indicate  that  the  lepton numbers $l_{e},
l_{\mu }, l_{\tau}$  are   well conserved, i.e. the standard weak
interactions do  not  violate  the lepton numbers.
\par
2) Then, in order to violate the lepton  numbers, it  is
necessary  to introduce an interaction violating these numbers. It
is equivalent to introducing nonseasonal  mass terms  in the  mass
matrix  of $\nu_{e}, \nu_{\mu }, \nu_{\tau}$. By diagonalizing
this matrix, we go to the $\nu_{1}, \nu _{2}, \nu_{3}$ neutrino
states. Exactly like it was in the case  of $K^{o}$  mesons
created  in strong interactions, when mainly $K^{o}, \bar{K}^{o}$
mesons were produced, in  the considered case $\nu_{e}, \nu_{\mu
}, \nu_{\tau}$, but not $\nu_{1}, \nu_{2}, \nu_{3}$, neutrino
states are mainly created in the weak interactions (this is so,
because the contribution of the lepton numbers violating
interactions  in this process is too small). And in such case no
oscillations take place.
\par
3) Then, when the $\nu_{e}, \nu_{\mu }, \nu_{\tau}$  neutrinos are
passing through vacuum, they  will  be  converted  into
superpositions  of  the $\nu_{1}, \nu _{2}, \nu_{3}$  owing  to
the presence  of  the interactions violating  the  lepton number
of neutrinos and  will be left on  their mass   shells.  And,
then, oscillations of the $\nu_{e}, \nu_{\mu}, \nu_{\tau}$
neutrinos will  take  place according to the standard scheme
[16-18]. Whether these oscillations are real or virtual, it will
be determined by the masses of the  physical neutrinos $\nu_{e},
\nu_{\mu}, \nu_{\tau}$.
\par
i) If the masses of the $\nu_{e}, \nu_{\mu }, \nu_{\tau}$
neutrinos  are equal, then the real oscillation of the neutrinos
will take  place.
\par
ii) If  the masses  of  the $\nu_{e}, \nu _{\mu }, \nu _{\tau}$
are  not equal, then the virtual oscillation of  the  neutrinos
will  take place. To make these oscillations  real,  these
neutrinos must participate  in the quasielastic interactions, in
order to undergo transition  to  the mass shell of the other
appropriate neutrinos by analogy with $\gamma  - \rho ^{o}$
transition  in the  vector   meson  dominance model.\\

{\bf 2. Neutrino Oscillation Types}\\

\par
The mass matrix of $\nu_e$ and $\nu_\mu$ neutrinos has the form
$$
\left(\begin{array}{cc} m_{\nu_e}& 0 \\ 0 & m_{\nu_\mu}
\end{array} \right) .
\eqno(1)
$$
\par
Due to the presence of the interaction violating the lepton
numbers, a nondiagonal term appears in this matrix and then this
mass matrix is transformed into the following nondiagonal matrix
($CP$ is conserved):
$$
\left(\begin{array}{cc}m_{\nu_e} & m_{\nu_e \nu_\mu} \\ m_{\nu_\mu
\nu_e} & m_{\nu_\mu} \end{array} \right) ,
\eqno(2)
$$
then the lagrangian of mass of the neutrinos takes the following
form ($\nu \equiv \nu_L$):
$$
\begin{array}{c}{\cal L}_{M} = - \frac{1}{2} \left[m_{\nu_e}
\bar \nu_e \nu_e + m_{\nu_\mu} \bar \nu_{\mu} \nu_{\mu } +
m_{\nu_e \nu_{\mu }}(\bar \nu_e \nu_{\mu } + \bar \nu_{\mu }
\nu _e) \right] \equiv \\
\equiv  - \frac{1}{2} (\bar \nu_e, \bar \nu_\mu)
\left(\begin{array}{cc} m_{\nu_e} & m_{\nu_e \nu_{\mu }} \\
m_{\nu_{\mu} \nu_e} & m_{\nu_\mu} \end{array} \right)
\left(\begin{array}{c} \nu_e \\ \nu_{\mu } \end{array} \right)
\end{array} ,
\eqno(3)
$$
which is diagonalized by turning through  the angle $\theta$ and
(see ref. in [16]) and then this lagrangian (3) transforms into
the following one:
$$
{\cal L}_{M} = - \frac{1}{2} \left[ m_{1} \bar \nu_{1} \nu_{1} +
m_{2} \bar \nu_{2} \nu_{2} \right]  ,
\eqno(4)
$$
where
$$
m_{1, 2} = {1\over 2} \left[ (m_{\nu_e} + m_{\nu_\mu}) \pm
\left((m_{\nu_e} - m_{\nu_\mu})^2 + 4 m^{2}_{\nu_\mu \nu_e}
\right)^{1/2} \right] ,
$$
\par
\noindent
and angle $\theta $ is determined by the following
expression:
$$
tg 2 \theta  = \frac{2 m_{\nu_e \nu_\mu}} {(m_{\nu_\mu} -
m_{\nu_e})} ,
\eqno(5)
$$
$$
\begin{array}{c}
\nu_e = cos \theta  \nu_{1} + sin \theta \nu_{2}  ,         \\
\nu _{\mu } = - sin \theta  \nu_{1} + cos \theta  \nu_{2} .
\end{array}
\eqno(6)
$$
From eq.(5) one can see that if $m_{\nu_e} = m_{\nu_{\mu}}$, then
the mixing angle is equal to $\pi /4$ independently of the value
of $m_{\nu_e \nu_\mu}$:
$$
sin^2 2\theta = \frac{(2m_{\nu_{e} \nu_{\mu}})^2} {(m_{\nu_e} -
m_{\nu_\mu})^2 +(2m_{\nu_e \nu_{\mu}})^2} ,
\eqno(7)
$$
$$
\left(\begin{array}{cc} m_{\nu_1} & 0 \\
0 & m_{\nu_2} \end{array} \right) .
$$

\par
It is interesting to remark that expression (7) can be obtained
from the Breit-Wigner distribution [19]
$$
P \sim \frac{(\Gamma/2)^2}{(E - E_0)^2 + (\Gamma/2)^2}   ,
\eqno(8)
$$
by using the following substitutions:
$$
E = m_{\nu_e},\hspace{0.2cm} E_0 = m_{\nu_\mu},\hspace{0.2cm}
\Gamma/2 = 2m_{\nu_e, \nu_\mu} ,
$$
where $\Gamma/2 \equiv W(... )$ is a width of $\nu_e \rightarrow
\nu_\mu$ transition, then we can use a standard method [18, 20]
for calculating this value.
\par
The expression for time evolution of $\nu _{1}, \nu _{2}$
neutrinos (see (4), (6)) with masses $m_{1}$ and $m_{2}$ is
\par
$$
\nu _{1}(t) = e^{-i E_1 t} \nu _{1}(0),  \qquad \nu _{2}(t) =
e^{-i E_2 t} \nu _{2}(0) ,
\eqno(9)
$$
where
$$
E^2_{k} = (p^{2} + m^2_{k}), k = 1, 2 .
$$
\par
If neutrinos are propagating without interactions, then
\par
$$
\begin{array}{c}
\nu_e(t) = cos \theta e^{-i E_1 t} \nu_{1}(0) + sin \theta
e^{-i E_2 t} \nu_{2}(0) , \\
\nu_{\mu }(t) = - sin \theta e^{-i E_1 t} \nu_{1}(0) + cos \theta
e^{-i E_2 t} \nu_{2}(0) .
\end{array}
\eqno(10)
$$
\noindent
Using the expression for $\nu _{1}$ and $\nu _{2}$  from
(6), and putting it into (10), one can get the following
expression:
$$
\nu_e (t) = \left[e^{-i E_1 t} cos^{2} \theta + e^{-i E_2 t}
sin^{2} \theta \right] \nu _e (0) +
$$
$$
+ \left[e^{-i E_1 t} - e^{-i E_2 t} \right] sin \theta \cos \theta
\nu_{\mu }(0) ,
\eqno(11)
$$
$$
\nu_{\mu }(t) = \left[e^{-i E_1 t} sin^{2} \theta + e^{-i E_2 t}
cos^{2} \theta \right] \nu_{\mu}(0)  +
$$
$$
+ \left[e^{-i E_1 t} - e^{-i E_2 t} \right] sin\theta cos \theta
\nu_e (0) .
$$
\par
The probability that neutrino $\nu_e$ created at the time $t = 0$
will be transformed into $\nu_{\mu}$ at the time $t$ is an
absolute value of amplitude $\nu_{\mu}(0)$ in (11) squared, i. e.
\par
$$
\begin{array}{c}
P(\nu_e \rightarrow \nu_{\mu}) = \mid(\nu_{\mu}(0) \cdot \nu_e(t)) \mid^2 =\\
 = {1\over 2} \sin^{2} 2\theta \left[1 - cos ((m^{2}_{2} - m^{2}_{1}) / 2p)
t \right] ,
\end{array}
\eqno(12)
$$
\noindent where it is supposed that $p \gg  m_{1}, m_{2}; E_{k}
\simeq p + m^{2}_{k} / 2p$.
\par
The expression (12) presents the probability of neutrino aroma
oscillations. The angle $\theta$ (mixing angle) characterizes
value of mixing. The probability $P(\nu_e \rightarrow  \nu_{\mu})$
is a periodical function of distances, where the period is
determined by the following expression:
$$
L_{o} = 2\pi  {2p \over {\mid m^{2}_{2} - m^{2}_{1} \mid}} .
\eqno(13)
$$
\par
And probability $P(\nu _e \rightarrow  \nu _e)$ that the neutrino
$\nu_e$ created at time $t = 0$ is preserved as $\nu_e$ neutrino
at time $t$ is given by the absolute value of the amplitude of
$\nu_e(0)$  in (11) squared. Since the states in (11) are
normalized states, then
$$
P(\nu_e \rightarrow  \nu_e) + P(\nu_e \rightarrow \nu_{\mu}) = 1 .
\eqno(14)
$$
\par
So, we see that aromatic oscillations caused by nondiagonality of
the neutrinos mass matrix violate the law of the $-\ell_e$ and
$\ell_{\mu}$ lepton number conservations. However, in this case, as
one can see from exp. (14), the full lepton numbers $\ell  =
\ell_e + \ell_{\mu}$ are conserved.
\par
We can also see that there are two cases of $\nu_e, \nu_\mu$
transitions (oscillations) [18], [20].\\

\par
1. If we consider the transition of $\nu_e$ into $\nu_\mu$
particle, then
$$
sin^2 2\beta \cong \frac{4m^2_{\nu_e, \nu_\mu}}{(m_{\nu_e} -
m_{\nu_\mu})^2 + 4m^2_{\nu_e, \nu_\mu}}  ,
\eqno(15)
$$
\par
How can we understand this  $\nu_e \rightarrow \nu_\mu$
transition?
\par
If $2m_{\nu_e, \nu_\mu} = \frac{\Gamma}{2}$ is not zero, then it
means that the mean mass of $\nu_e$ particle is $m_{\nu_e}$ and
this mass is distributed by $sin^2 2\beta$ (or by the Breit-Wigner
formula) and the probability of the $\nu_e \rightarrow \nu_\mu$
transition differs from zero and it is defined by masses of
$\nu_e$ and $\nu_\mu$ particles and $m_{\nu_e, \nu_\mu}$, which is
computed in the framework of the standard method, as pointed out
above.
\par
So, this is a solution of the problem of the origin of mixing
angle in the theory of vacuum oscillations.
\par
In this case the probability of $\nu_e \rightarrow \nu_\mu$
transition (oscillation) is described by the following expression:
$$
P(\nu_e \rightarrow \nu_\mu, t) =  sin^2 2\beta sin^2 \left[\pi
t\frac{\mid m_{\nu_1}^2 - m_{\nu_2}^2 \mid}{2 p_{\nu_e}} \right ],
\eqno(16)
$$
where $p_{\nu_e}$ is a momentum of $\nu_e$ particle.
\par
Originally it was supposed [6, 16] that these oscillations are
real oscillations, i.e. that there takes place real transition of
electron neutrino $\nu_e$ into muon neutrino $\nu_{\mu}$ (or tau
neutrino $\nu_{\tau}$). Then the neutrino $x = \mu, \tau$ is
decayed in electron neutrino plus something
$$
\nu_{x} \rightarrow \nu_e + ....  ,
\eqno(17)
$$
as a result we get energy from vacuum, which equals the mass
difference (if $m_{\nu_x} > m_{\nu_e}$)
$$
\Delta E \sim m_{\nu_{x}} - m_{\nu_e} .
\eqno(18)
$$
Then, again this electron neutrino transits into muon  neutrino,
which is decayed again and we get energy and etc. {\bf So we got a
perpetuum mobile!} Obviously, the law of energy conservation cannot
be fulfilled in this process. The only way to restore the law of
energy conservation is to demand that this process is virtually one.
Then, these oscillations will be the virtual ones and they are
described in the framework of the uncertainty relations. The correct theory
of neutrino oscillations can be constructed only  into the
framework of the particle physics theory, where the conception
of mass shell is [17], [18], [20].\\
\par
2. If we consider the virtual transition of $\nu_e$ into $\nu_\mu$
neutrino at  $m_{\nu_e} = m_{\nu_\mu}$ (i.e. without changing the
mass shell), then
$$
tg 2\beta = \infty  ,
\eqno(19)
$$
$\beta = \pi/4$, and
$$
sin^2 2\beta = 1     .
\eqno(20)
$$
\par
In this case the probability of the $\nu_e \rightarrow \nu_\mu$
transition (oscillation) is described by the following expression:
$$
P(\nu_e \rightarrow \nu_\mu, t) = \left[\pi t\frac{4 m_{\nu_e,
\nu_\mu}^2}{2 p_a} \right ] .
\eqno(21)
$$
\par
In order to make these virtual oscillations real, their participation in
quasielastic interactions is necessary for the transitions to
their own mass shells [20].
\par
It is clear that the $\nu_e \rightarrow \nu_\mu$ transition is a
dynamical process.
\par
The question is: which type of neutrino oscillations is realized
in the Nature?\\

\par
3. The third type of transitions (oscillations) can be realized by mixings of
the fields (neutrinos) in analogy with the  vector  dominance
model ($\gamma-\rho^o$ and $Z^o-\gamma$  mixings) in a way as it takes place
in the particle physics. Since the weak couple constants
$g_{\nu_e}, g_{\nu_{\mu}},  g_{\nu_{\tau}}$ of $\nu_e, \nu_{\mu},
\nu_{\tau}$  neutrinos nearly are equally in reality, i.e.
$g_{\nu_e}\simeq g_{\nu_{\mu}} \simeq g_{\nu_{\tau}}$  then the
angle mixings are nearly maximal:
$$
sin \theta_{\nu_{e} \nu_{\mu}} \simeq
\frac{g_{\nu_e}}{\sqrt{g^2_{\nu_{e}} + g^2_{\nu_{\mu}}}}
=\frac{1}{\sqrt{2}} \simeq
sin \theta_{\nu_{e} \nu_{\tau}} \simeq sin \theta_{\nu_{\mu}
\nu_{\tau}}.
\eqno(22)
$$
Therefore, if the masses of these neutrinos are  equal (which is
hardly probable), then  transitions between neutrinos will be real
and if the masses of these  neutrinos are not equal then transitions
between  neutrinos will be virtual in analogy with $\gamma-\rho^o$
transitions.\\

{\bf 3. Mechanisms of Neutrino oscillations}\\

\par
{\bf 3.1. Impossibility of resonance enhancement of neutrino oscillations in
matter}\\

\par
In three different approaches: by using mass Lagrangian [10,21,
22], by using the Dirac equation [21, 22], and using the operator
formalism [23], the author of this work has discussed the problem
of the mass generation in the standard weak interactions and has
come to a conclusion that the standard weak interaction cannot
generate masses of fermions since the right-handed components of
fermions do not participate in these interactions. It is also
shown [24] that the equation for Green function of the
weak-interacting fermions (neutrinos) in the matter coincides with
the equation for Green function of fermions in vacuum and the law
of conservation of the energy and the momentum of neutrino in
matter will be fulfilled [23] only if the energy $W$ of
polarization of matter by the neutrino or the corresponding term
in Wolfenstein equation, is zero (it means that neutrinos cannot
generate permanent polarization of matter).  These results lead to
the conclusion:  resonance enhancement of neutrino oscillations in
matter does not exist.
\par
The simplest method to prove the absence of the resonance
enhancement of neutrino
oscillations in matter is:
\par
   If we put an electrical($e$) (or strong ($g$)) charged particle $a$ in  vacuum,
there arises polarization of vacuum. Since the field around particle $a$
is spherically symmetrical, the polarization must also be spherically
symmetrical. Then the particle will be left at rest and the law of
energy and momentum conservation is fulfilled.
\par
If we put a weakly ($g_W$) interacting particle $b$ (a neutrino)
in vacuum, then since the field around the particle has a
left-right asymmetry (weak interactions are left interactions with
respect to the spin direction [25, 29]), polarization of vacuum
must be nonsymmetrical, i.e. on the left side there arises maximal
polarization and on the right there is zero polarization. Since
polarization of the vacuum is asymmetrical, there arises
asymmetrical interaction of the particle (the neutrino) with
vacuum and the particle cannot be at rest and will be accelerated.
Then the law of energy momentum conservation will be violated. The
only way to fulfill the law of energy and momentum conservation is
to demand that polarization of vacuum be absent in the weak
interactions. The same situation will take place in matter. It is
necessary to remark that for above considered proof it is sufficient to know
that the field around of the weakly interacting particle is
asymmetrical (and there is no a necessary to know the  precise
form of this field). It is necessary also to remark that the
Super-Kamiokande datum on day-night asymmetry [13] is
$$
A = (D - N)/(\frac{1}{2}(D + N)) = -0.021 \pm 0.020(stat) +
\eqno(23)
$$
$$
+ 0.013(-0.012)(syst) .
$$
and it does not leave hope on possibility of the resonance
enhancement of neutrino oscillations in matter.
\par
In means that the forward scattering amplitude of the weak
interactions have a specific behavior.
\par
It is interesting to remark that in the gravitational interaction
the polarization does not exist either.\\

\par
{\bf An Amplitude of forward scattering at the weak
\par
interactions}\\

\par
Connection between refraction coefficient $n$ and amplitude of forward
scattering $f_k (k, 0)$ in matter is given by the following expression:
$$
n -1 = \frac{2 \pi}{k^2} \sum_i Re f_i (k, 0) ,
\eqno(24)
$$
where $i$ is index of summation and $k, p$ respectively are momentum and
transfer momentum.
\par
Since at the weak interactions the polarization is absent, then
$$
n = 1 ,
\eqno(25)
$$
and
$$
Re f_i (k, 0) = 0 ,
\eqno(26)
$$
i.e.
$$
Re f_i (k, p) \rightarrow 0 ,
\eqno(27)
$$
$$
p \rightarrow 0 .
$$
The amplitude of forward scattering goes to zero when transfer
momentum goes to zero in contrast to the strong and
electromagnetic interactions, where it differs from zero.\\

\par
{\bf 3.2. Majorana Neutrino Oscillations}\\
\par
At present it is supposed [26]
that the neutrino oscillations can be connected with
Majorana neutrino oscillations. I will show that we cannot put
Majorana neutrinos in the standard Dirac theory. It means that on
experiments the Majorana neutrino oscillations cannot be observed.
\par
Majorana fermion in Dirac representation has the following form
[6, 16, 27]:
\par
$$
\chi^M = \frac{1}{2} [\Psi(x) + \eta_C\Psi^{C}(x)] ,
\eqno(28)
$$
$$
\Psi^C(x) \rightarrow \eta_C C \bar \Psi^T(x) ,
$$
\par
\noindent where $\eta_{C}$ is a phase, $C$  is a charge
conjunction, $T$ is a transposition.
\par
From Exp. (28) we see that Majorana fermion $\chi^M$ has two spin
projections $\pm \frac{1}{2}$ and then the Majorana spinor can be
rewritten in the following form:
\par
$$
\chi^M (x) = \left(\begin{array}{c} \chi_{+\frac{1}{2}}(x)\\
\chi_{-\frac{1}{2}}(x) \end{array} \right) .
\eqno(29)
$$
The mass Lagrangian of Majorana neutrinos in the case of two
neutrinos $\chi_e, \chi_\mu$ ($-\frac{1}{2}$ components of
Majorana neutrinos, and $\bar \chi_{...}$ is the same Majorana
fermion with the opposite spin projection) in the common case has
the following form:
$$
\begin{array}{c} { \cal L}^{'}_{M} =
 - \frac{1}{2}(\bar \chi_e, \bar \chi_\mu)
\left(\begin{array}{cc} m_{\chi_e} & m_{\chi_e \chi_\mu} \\
m_{\chi_\mu \chi_e} & m_{\chi_\mu} \end{array} \right)
\left(\begin{array}{c} \chi_e \\ \chi_\mu \end{array} \right)
\end{array} .
\eqno(30)
$$
Diagonalizing this mass matrix by standard methods one obtains the
following expression:
$$
\begin{array}{c}  {\cal L}^{'}_{M} =
 - \frac{1}{2}(\bar \nu_1, \bar \nu_2)
\left(\begin{array}{cc} m_{\nu_1} & 0 \\
0 & m_{\nu_2} \end{array} \right) \left(\begin{array}{c} \nu_1 \\
\nu_2 \end{array} \right) \end{array} ,
\eqno(31)
$$
where
$$
\nu_1 = cos \theta \chi_e - sin \theta \chi_\mu ,
$$
$$
\nu_2 = sin \theta \chi_e + cos \theta \chi_\mu   .
$$
These neutrino oscillations are described by expressions (9)-(14)
with the following substitution of $ \nu_{e \mu}  \to \chi^M_{e
\mu}$ .
\par
The standard theory of weak interactions is constructed on the
base of local gauge invariance of Dirac fermions. In this case
Dirac fermions have the following lepton numbers $l_{l,}$ which
are conserved,
\par
$$
l_{l}, l = e ,\mu , \tau,
\eqno(32)
$$
\noindent
and Dirac antiparticles have lepton numbers with the
opposite sign
\par
$$
\bar{l} = - l_{l}.
\eqno(33)
$$
\par
Gauge transformation of Majorana fermions can be written in the
form:
$$
{\chi'}_{+\frac{1}{2}}(x) = exp(-i\beta) \chi_{+\frac{1}{2}}(x) ,
$$
$$
{\chi'}_{-\frac{1}{2}}(x) = exp(+i\beta) \chi_{-\frac{1}{2}}(x)  .
\eqno(34)
$$
Then lepton numbers of Majorana fermions are
\par
$$
l^{M} =\sum_{i} l^{M}_{i} (+1/2) = -\sum_{i} l^{M}_{i}(-1/2) ,
$$
\noindent
i. e., antiparticle of Majorana fermion is the same
fermion with the opposite spin projection.
\par
Now we come to discussion of the problem of the place of Majorana
fermion in the standard theory of weak interactions [28].
\par
To construct the standard theory of weak interactions [29], Dirac
fermions are used. The absence of contradiction of this theory
with the experimental data confirms that all fermions are Dirac
particles.
\par
Now, if we want  to put the Majorana fermions into the standard
theory we must take into account that, in the common case, the
gauge charges of the Dirac and Majorana fermions are different
(especially it is well seen in the example of Dirac fermion having
an electrical charge since it cannot have a Majorana charge (it is
worth to remind that in the weak currents the fermions are
included in the couples form)). In this case we cannot just
include Majorana fermions in the standard  theory of weak
interactions by gauge invariance manner. Then, in the standard
theory the Majorana fermions cannot
appear. \\

\par
{\bf 3.3. Transitions (Oscillations) of Aromatic Neutrinos}\\
\par
 In the work [2] Maki et al. supposed that there  could exist
transitions between aromatic neutrinos $\nu_e, \nu_\mu$.
Afterwards $\nu_\tau$ was found and then $\nu_e, \nu_\mu,
\nu_\tau$ transitions could be possible. The author of this work
has developed this direction (see [21, 25]). It is necessary to remark
that only this scheme of oscillations is realistic for neutrino
oscillations (see also this work). The expressions, which described
neutrino oscillations in this case are given above
in expressions (9)-(14).\\

\par
{\bf III. EXPERIMENTS}\\

\par
{\bf 1. Experimental Observation of the  Neutrino Oscillations}\\

\par
 At present, it is supposed that the neutrino oscillations were
observed [13-15]. In these experiments only transitions between
the Sun or atmospheric neutrinos have been observed. Since we
suppose that there take place the neutrino oscillations, therefore
we must observe (Sun) neutrino oscillations in reality. Since the
length of neutrino oscillations is sufficiently great we cannot
observe higher  modes in terrestrial experiment. But we have
another possibility to observe the Sun neutrino oscillation using
the fact that the Earth orbit is elliptic one with:\\

\par
Earth's perihelion $R_P$ = 147.117 $\cdot 10^6 km$ ,
\par
Earth's aphelion $R_A$ = 152.083 $\cdot 10^6 km$,
\par
\noindent
and their difference $\Delta R$ is $\Delta R = 4.866 \cdot 10^6 km$.
Since the Sun neutrinos conclude all energies up to $15 MeV$, we
must divide this energy spectrum on energy regions (also distances
must be divided on regions) and observe these neutrino fluxes as
function of energy and the Ears's distances from the Sun. At these conditions
we must observe the neutrino oscillations if the length of
neutrino oscillation $R_{osc}$ is bigger than the region $\Delta$,
where these (high energy) neutrinos are generated on the Sun i.e.
$$
\Delta \sim 0.05 R_{sun} \sim 10^4 km ,
\eqno(35)
$$
$$
R_{osc} > \Delta .
$$
It is obvious that in these experiments it is impossible to register
$\nu_{\mu} \leftrightarrow \nu_{\tau}$ oscillations since their
length, as we know from Super-Kamiokande experiments, is  small enough.
The Super-Kamiokande and SNO detectors wholly fit for such
observations.\\

\par
{\bf 2. The Earth Neutrinos}

At present, there exist big detectors on the  Sun  and  atmospheric
neutrinos observation. The problem study  of the Earth neutrino
sources present enormous  interest, therefore, using  the same
detectors it is possible to research the Earth neutrino sources.
More detailed consideration this question will  be published later. It is
important that there is no a necessity to reconstruct the detectors.
It is only necessary to collect data from the Earth as it is
fulfilled for  the Sun neutrinos but for opposite to the Sun
direction [30]. In this way we can obtain the Earth neutrino sources map
using  the detectors located in different  places of the Earth  surface. \\
\vspace{1cm}

{\bf 3. The Problem with GNO (GALLEX), SAGE Data}\\

\par
As stressed above, transitions between neutrinos with different
flavor have been already observed and the neutrino mixing angles are
nearly maximal. The  Super-Kamoikande [13], SNO [15], Homestake
[31] data (D) normalized on SSM calculations (see [32)) are in
good agreement.
\par
Homestake 1970-1994, $E_{thre} = 0.814$:
$$
 \nu_e + ^{71}Ga \to ^{71}Ge  + e^{-},  \qquad
 \frac{D^{exp}}{D^{BP2000}} = 0.34 \pm 0.03
$$
\par
Super-Kamiokande 1996-2001, $E_{thre} = 4.75 MeV$:
$$
\nu_e  +  e^{-} \to \nu_e + e^{-}, \qquad
\frac{D^{exp}}{D^{BP2000}} = 0.465 \pm 0.015 ;
$$
\par
SNO $E_{thre} = 6.9 MeV$
$$
\nu_e + d \to  p + p + e^{-}, \qquad \frac{D^{exp}}{D^{BP2000}} =
0.35 \pm 0.02,
$$
\par
$E_{thre} = 2.2 MeV$
$$
\nu_e + d \to  p + n + e^{-}, \qquad \frac{D^{exp}}{D^{BP200}} =
1.01 \pm 0.13,
$$
\par
$E_{thre} = 5.2 MeV$
$$
\nu + e^{-} \to  \nu + e^{-}, \qquad \frac{D^{exp}}{D^{BP200}} =
0.47 \pm 0.05,
$$
But normalized on SSM [32] GNO (GALLEX) [33], SAGE [12, 34]
data are higher than above data on
values $0.16-0.20$
\par
GNO (GALLEX) 1998-2000, $E_{thre} = 0.233 MeV$
$$
 \nu_e + ^{71}Ga \to ^{71}Ge  + e^{-},  \qquad
 \frac{D^{exp}}{D^{BP200}} = 0.51 \pm 0.08;
$$
\par
SAGE 1990-2001, $E_{thre} = 0.233 MeV$
$$
 \nu_e + ^{71}Ga \to ^{71}Ge  + e^{-},  \qquad
 \frac{D^{exp}}{D^{BP200}} = 0.54 \pm 0.05;
$$
\par
Since the length of neutrino transitions (oscillations) is
proportional to their energy (see eq.(13)), therefore,
the smaller are the neutrino energies, the smaller are the lengths of transitions (oscillations),
hence the norm of  transitions (oscillations) at
the small energies must be the same as at the high ones.
In order to resolve this problem it is probably necessary to examine calibration of the
last experiments. It is more likely that the Standard Sun model [32]
requires a revision in these energy regions.  It is very important
to  organize an experiment with neutral currents in this energy
region in order to register the full (the Sun) neutrino fluxes.
It is clear that there is no sense in drawing the allowed regions pictures until this problem is
solved.  \\

\par
{\bf 4.The CHOOZ problem}
\par
In the France reactor experiment on $\bar \nu_e$ neutrinos, there was
observed a very small angle mixing for $\bar \nu_e \rightarrow \bar \nu_{\tau}$
transitions [35]. If it is correct then on  SNO there should be
observed [15] only the $\nu_e \rightarrow \nu_{\mu}$ neutrino transitions
and the $\nu_e \rightarrow \nu_{\tau}$ transitions must be suppressed, i.e.
relation between $\nu_e $ and $\nu_{\mu}$ neutrino fluxes must be
equal
$$
\Phi_{\nu_e} \simeq \Phi_{\nu_{\tau}} .
\eqno(36)
$$
However, in the SNO experiments on neutral currents, we can see
approximate equality the numbers of the three type of neutrinos.
$$
\Phi_{\nu_e} \simeq \frac{1}{2}(\Phi_{\nu_{\tau}} +
\Phi_{\nu_{\mu}})    .
\eqno(37)
$$
It is clear that the angle of $\bar \nu_e \rightarrow \bar
\nu_{\tau}$ transition cannot be small. Since distance from the reactor
is small then this detector can register only a small mixing angle (to see
correct mixing angle the detector must be in a distance not smaller than
the oscillations length).  \\

{\bf Conclusion}\\

It is noted that the theory of neutrino oscillations can be
constructed only in the framework of the particle physics theory,
where is a mass shell conception and then transitions
(oscillations) between neutrinos with equal masses are real  and
between neutrinos  with different masses are virtual.
\par
It is necessary to solve the question: which type of neutrino
transitions (oscillations) is realized in nature? There can be
three  types of neutrino transitions (oscillations).
\par
At present  it is considered that Dirac and Majorana neutrino
oscillations can be realized. It is shown that we cannot put
Majorana neutrinos in the standard weak interactions theory without
violation of  the gauge invariance. Then it is obvious that there
can be only realized transitions (oscillations) between Dirac
neutrinos with different flowers.
\par
Also it is shown that the mechanism of
resonance enhancement of neutrino oscillations in matter cannot be
realized without violation of the law of energy-momentum
conservation.
\par
Though it is supposed that we see neutrino oscillations
in experiments, indeed there only transitions between neutrinos are registered.
In order to register neutrino oscillations it  is
necessary to see second or even higher neutrino oscillation modes
in experiments. For this purpose we can use the elliptic character
of the Earth orbit.
\par
The analysis shows that the SNO
experimental results do not confirm smallest of $\nu_e \to
\nu_{\tau}$ transition angle mixings, which was obtained in CHOOZ
experiment. It is also noted that there is contradiction between SNO,
Super-Kamiokande, Homestake and the SAGE and GNO (GALLEX) data.\\

\par
{\bf References}\\

\par
\noindent 1. Pontecorvo B. M., Soviet Journ. JETP, 1957, v. 33,
p.549;
\par
JETP, 1958,  v.34, p.247.
\par
\noindent 2. Maki Z. et al., Prog.Theor. Phys., 1962, vol.28,
p.870.
\par
\noindent 3. Pontecorvo B. M., Soviet Journ. JETP, 1967, v. 53,
p.1717.
\par
\noindent 4.
Davis R. et al., Phys. Rev. Letters, 1968, vol.20,
p.1205.
\par
\noindent 5.
Bahcall J. et al., Phys. Lett.B, 1968, vol.26, p.1;
\par
Bahcall J.,  Bahcall N., Shaviv  G.,  Phys.  Rev.  Lett.  1968,
vol.20,
\par
p.1209;
\par
   S. Turck-Chieze et al., Astrophys.J. 335 (1988), p.415.
\par
\noindent 6.
Gribov V., Pontecorvo B.M., Phys. Lett. B, 1969,
vol.28, p.493.
\par
\noindent 7.
Hirata K.S. et  al., Phys. Rev. Lett., 1989,
vol.63,p.16.
\par
\noindent 8.
Mikheyev S.P., Smirnov A.Ju., Nuovo Cimento, 1986,
vol.9,p.17.
\par
\noindent 9.
Wolfenstein L., Phys. Rev.D, 1978, vol.17, p.2369.
\par
\noindent 10.
Beshtoev Kh.M., JINR Commun. E2-91-183, Dubna, 1991;
\par
Proceedings of III  Int. Symp. on Weak and Electromag. Int. in
\par
Nucl. (World Scient., Singapoure, P. 781, 1992);
\par
13th European Cosmic Ray Symp. CERN, Geneva, HE-5-13.
\par
\noindent 11.
Anselmann P. et al., Phys. Lett. B, 1992,
\par
vol.285, p.376;  1992, vol.285, p.391;
\par
Hampel W. et al., Phys. Lett. B, 1999, v.447, p. 127.
\par
\noindent 12.
Abdurashitov  J.N.  et al.,  Phys.  Lett.B,  1994,
vol.328,
\par
p.234; Phys. Rev. Lett., 1999, v.83,  p.4683.
\par
\noindent
13. Kameda J., Proceedings of ICRC 2001, August 2001,
Germany,
\par
Hamburg, p.1057.
\par
Fukuda  S. et al,. Phys.   Rev . Lett. 2001, v. 25, p.5651;
\par
Phys. Lett. B 539, 2002,  p.179.
\par
\noindent
14. Toshito T., hep-ex/0105023;
\par
Kameda J., Proceeding of 27th ICRC, August 2001;
\par
Hamburg, Germany, v.2, p.1057.
\par
Mauger Ch., 31-st ICHEP, Amsterdam, July 2002.
\par
\noindent
15. Ahmad Q. R. et al., Internet Pub. nucl-ex/0106015,
June 2001.
\par
Ahmad  Q. R. et al., Phys. Rev. Lett. 2002, v. 89, p.011301-1;
\par
Phys. Rev. Lett.  2002,v.  89, p.011302-1.
\par
\noindent
16. Bilenky S.M., Pontecorvo B.M., Phys. Rep., C41(1978)225;
\par
Boehm F., Vogel P., Physics of Massive Neutrinos: Cambridge
\par
Univ. Press, 1987, p.27, p.121;
\par
Bilenky S.M., Petcov S.T., Rev. of Mod.  Phys., 1977, v.59,
\par
p.631.
\par
\noindent
17. Beshtoev Kh.M., JINR Commun. E2-92-318, Dubna, 1992;
\par
JINR Rapid Communications, N3[71]-95.
\par
\noindent
18. Beshtoev Kh.M., Internet Pub. hep-ph/9911513;
\par
 The Hadronic Journal, v.23, 2000, p.477;
\par
Proceedings of 27th Intern. Cosmic Ray Conf., Germany,
\par
Hamburg, 7-15 August 2001, v.3, p. 1186.
\par
\noindent
19. Blatt J.M., Waiscopff V.F., The Theory of Nuclear Reactions,
\par
INR T.R. 42.
\par
\noindent
20. Beshtoev Kh.M., JINR Commun. E2-99-307, Dubna, 1999;
\par
JINR Commun. E2-99-306, Dubna, 1999.
\par
\noindent
21. Beshtoev Kh.M., Phys. of Elem. Part. and Atomic Nucl.
\par
(Particles and Nuclei), 1996, v.27, p.53.
\par
\noindent
22. Beshtoev Kh. M., JINR Communication E2-93-167, Dubna,
\par
1993; JINR Communication P2-93-44, Dubna, 1993;
\par
\noindent
23. Beshtoev Kh.M., HEP-PH/9912532, 1999;
\par
Hadronic Journal, 1999, v.22, p.235.
\par
\noindent
24. Beshtoev Kh.M., JINR Communication E2-2000-30, Dubna,
\par
2000; Internet Publ. hep-ph/0003274.
\par
\noindent
25. Beshtoev Kh.M., JINR Communication D2-2001-292, Dubna,
\par
2001; Internet Publ. hep-ph/01        03274.
\par
\noindent
26. C. Gonzalez-Garcia 31-st ICHEP, Amsterdam, July 2002.
\par
\noindent
27. Rosen S.P., Lectore Notes on Mass Matrices, LASL preprint,
\par
1983.
\par
\noindent
28. Beshtoev Kh.M., JINR Commun. E2-92-195, Dubna, 1992.
\par
\noindent
29. Glashow S.L.- Nucl. Phys., 1961, vol.22, p.579 ;
\par
Weinberg S.- Phys.  Rev. Lett., 1967, vol.19, p.1264 ;
\par
Salam A.- Proc. of the 8th Nobel  Symp.,  edited  by
\par
N. Svarthholm (Almgvist and Wiksell,  Stockholm) 1968, p.367.
\par
\noindent
30. Mitsui T., Proceedings of the 31-st ICHEP, August 2002,
\par
Amsterdam, Netherlands.
\par
\noindent
31. Davis R. et al., Phys. Rev. Letters, 1968, vol.20;
\par
Lande  K., Neutrino 2002, may 2002, Munich, Germany.
\par
\noindent
 32. Bahcall J., Pinsonneault M.N. and Sarbani B. Astrophysical
\par
 Journ 2001,   v. 555, p.990.
\par
\noindent
33. Belloti E.,  Nucl. Phys.B (Proc. Suppl.), 2001,v. 91,   p.44;
\par
Hampel W.  et al., Phys. Lett.B, 1996,  v. 388,  p.364;
\par
Kirsten T., Neutrino 2002, May 2002, Munich, Germany.
\par
\noindent
34. Gavrin V. N., Nucl. Phys. B (Proc. Suppl.) 2001, v. 91, p.36.
\par
\noindent
 35. Appolinio M., Phys. Lett. B, 1999, v.466, p.415.

\end{document}